\documentclass[aps,prb,epsfigm,twocolumn,showpacs]{revtex4-1}
\usepackage{graphicx}
\usepackage{amsmath}

\begin{document}

\title{Emission spectrum of quasi-resonant laterally coupled quantum dots}

\author{Miquel Royo, Juan I. Climente, Josep Planelles}
\affiliation{Departament de Qu\'{\i}mica F\'{\i}sica i Anal\'{\i}tica,
Universitat Jaume I, E-12080, Castell\'o, Spain}
\email{josep.planelles@uji.es}
\homepage{http://quimicaquantica.uji.es/}
\date{\today}

\begin{abstract}
We calculate the emission spectrum of neutral and charged excitons in a pair of
laterally coupled InGaAs quantum dots with nearly degenerate energy levels.
As the interdot distance decreases, a number of changes take place in the
emission spectrum which can be used as indications of molecular coupling.
These signatures ensue from the stronger tunnel-coupling of trions as compared 
to that of neutral excitons.
\end{abstract}


\maketitle

\section{Introduction}

There is current interest in developing quantum dot molecules (QDMs)
for quantum information processing.\cite{BurkardPRB,BayerSCI}
Great progress in this direction has been made using vertical QDMs 
fabricated with epitaxial growth techniques.  These structures are 
formed by pairs of vertically stacked QDs with a thin barrier in between.
Because the constituent QDs are generally asymmetric, electric
fields applied along the coupling direction are used to control
the charge and the tunnel coupling strength.\cite{OrtnerPRL,KrennerPRL,KrennerPRL2,StinaffSCI}

In the last years, developement of QDMs formed by laterally coupled QD
pairs is also being pursued.\cite{BairaJAP,WangNJP,BeirnePRL,AlonsoJCG,YamagiwaAPL,LeeIEEE}
In these structures, the QD pair is formed inside the same epitaxial layers.
This architecture is expected to offer some advantages as compared to
vertical QDMs.\cite{WangAM,KrapekJPCS} In particular, independent control
of QD charging and tunneling strength should be possible through the
application of vertical and lateral electric fields, respectively.
Also, simultaneous coupling of multiple QDMs, which is a requirement
for the scaling of qubit operations, should be feasible through the
implementation of individual electric gates for each QDM.
On the other hand, lateral QDMs present their own obstacles.
For example, the distance between the centers of the QDs is much
larger than in vertical structures, which implies weaker tunneling.
Indeed, despite recent advances in the electrical engineering of lateral QDMs,
which include the demonstration of Coulomb blockade charging\cite{MunozJPCS,AlenXXX,Zhou_arxiv}
and tuning of optical resonances\cite{HermannstadterPRB},
the application of lateral electric fields producing clear spectroscopic
signatures of molecular coupling remains a challenge.

Understanding the differences between the physics of lateral QDMs and
the better known case of vertical QDMs will contribute to determine the conditions 
for optimal device performance, be it vertical, lateral or hybrid.\cite{HatanoSCI}
Many of the differences arise from the distinct synthetic routes, 
which lead to characteristic structural and compositional profiles.\cite{HermannstadterPRB,PengPRB}
Thus, in lateral QDMs it has been shown that the weaker quantum confinement 
of the QDs along the coupling direction leads to an enhanced role of the excited 
orbitals within the QDs.\cite{SzafranPRB,BarticevicJPCM} Also, the nonradiative 
relaxation dynamics of lateral InGaAs QDMs has been shown to follow a singular path.\cite{PengPRB,PengPRB2}

In this paper, we study the optical resonances of neutral, negatively-charged and 
positively-charged excitons in lateral InGaAs QDMs as a function of the interdot distance. 
The dots have slightly different energies, as is often the case with current growth 
techniques.\cite{BairaJAP,BeirnePRL,MunozJPCS,AlenXXX,HermannstadterPRB}
We find that tunnel-coupling of trions is stronger than that of neutral excitons.
This leads to signatures in the photoluminescence spectrum at zero electric field
which can be used to distinguish uncoupled QD pairs from coupled QDMs.
The convenience of using trions instead of neutral excitons for quantum information 
protocols is discussed.

\section{Theoretical Model}

Our Hamiltonian for excitons and trions can be written in the second quantization as:

\begin{multline}
{\hat H} =  \sum_i E_i^e\,e_i^+\,e_i  +  \sum_p E_p^h\,h_p^+\,h_p  
+ \frac{1}{2} \sum_{ijkl} \langle ij| \,V\, | kl\rangle \, e_i^+  e_j^+  e_k  e_l\\
+ \frac{1}{2} \sum_{pqrs} \langle pq| \,V\, | rs\rangle \, h_p^+  h_q^+  h_r  h_s 
+ \sum_{ijpq} \langle ip| \,V\, | qj\rangle \, e_i^+  h_p^+  h_q  e_j ,
\label{Hfull}
\end{multline}
 
\noindent where $E_i^e$ ($E_p^h$) is the electron (hole) energy in the
single-particle state $|i\rangle$ ($|p\rangle$), $e_i^+/e_i$ ($h_p^+/h_p$) is the electron
(hole) creation/annihilation operator, i.e., ($e_i^+\,e_j^+\dots$ )($h_p^+\,h_q^+\dots$)
$|0_h\rangle |0_e\rangle=|p\,q\dots\rangle |i\,j\dots\rangle$, and $\langle ij|\,V\,| kl \rangle$,
$\langle pq|\,V\,| rs \rangle$ and $\langle ip| \,V\, | qj\rangle$ are the
electron-electron, hole-hole and electron-hole Coulomb matrix elements respectively. 
Electron-hole exchange term is safely neglected because in the absence of 
magnetic fields and spin mixing, it has no effect on the observed emission spectrum.

To calculate the single-particle states, we use a two-dimensional effective 
mass Hamiltonian for electrons and heavy holes:

\begin{equation}
{\hat H}_i = \frac{1}{2\,m_i^*} \, (p_x^2 + p_y^2) + V_i(x,y), 
\label{Hsp}
\end{equation}

\noindent where $i=e,h$ is the index denoting electrons or holes, 
$m^*$ is the effective mass, $p_{\alpha}=-i\hbar\nabla_{\alpha}$
the momentum operator, and $V(x,y)$ the quantum confinement potential.
For a pair of QDs whose centers are separated by a distance $d$, $V(x,y)$
is defined with intersecting parabolic potentials:

\begin{multline}
V_i(x,y)=\frac{1}{2}\,m_i^*\,
\left[ \min\left( \omega_{i,L}^2 \,((x + \frac{d}{2})^2 + y^2), \right. \right. \\
\left. \left. \omega_{i,R}^2  \,((x - \frac{d}{2})^2 + y^2) \right) \right].
\label{Vxy}
\end{multline}

\begin{figure}[h!]
\includegraphics[width=0.4\textwidth]{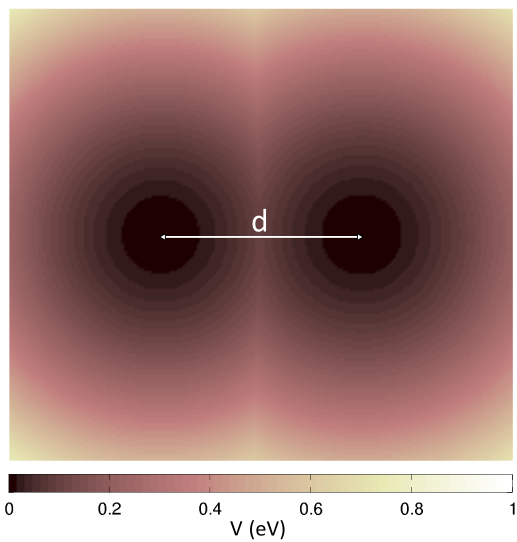}
\caption{(Color online). Contour plot of the confinement potential for
slightly asymmetric, laterally coupled QDs with an interdot distance d=40 nm.}
\label{fig:Vxy}
\end{figure}

\noindent Here $\omega_{i,L}$ ($\omega_{i,R}$) is the confinement frequency
in the left (right) QD.  A contour plot of this potential profile for a slightly
asymmetric QDM  is shown in Figure \ref{fig:Vxy}. Similar models were previously
adopted to study laterally coupled QDs~\cite{SzafranPRB05,HarjuPRL02}. 

Hamiltonian (\ref{Hsp}) is integrated numerically using a finite differences
scheme. The resulting electron and hole states are plugged into Hamiltonian
(\ref{Hfull}), together with the Coulomb matrix elements,
which we calculate using statistical methods. The Hamiltonian is then solved 
using a configuration interaction routine.\cite{citool} 
In this way, we obtain an accurate description of the ground and excited states 
of the QDM system. To simulate their emission spectrum, we use the
dipole approximation and Fermi's golden rule.\cite{Hawrylak_book}
The recombination probability from an initial few-body state $|i\rangle$
to a final state $|f\rangle$ with one less electron-hole pair, at an 
emission frequency $\omega$ is then given by:

\begin{equation}
\tau^{-1}_{f \leftarrow i} (\omega) \propto 
| \langle f|\, {\cal {\hat P}} \,| i \rangle |^2   
\; \Gamma(E^i - E^f -\hbar \omega) \; p_i(T),
\label{eq:emission}
\end{equation}

\noindent  where ${\cal {\hat P}}$ is the polarization operator\cite{RajadellJPCc},
$E^s$ is the energy of the state $|s\rangle$ and $\Gamma(E)$ is a Lorentzian curve centered at 
energy $E$, which simulates the intrinsic bandwidth of the transition. 
$p_i(T)$ is the thermal population distribution function for the initial state at temperature $T$.
We shall consider two possible dynamics for the emission process: (i) emission
takes place from all the excited states, which are populated through non-resonant 
excitation, prior to their thermal relaxation. We simulate this adopting the 
same occupation probability for all the initial states, $p_i(T)=1$;
(ii) recombination once the thermal equilibrium is reached. In this case we
assume a Boltzmann distribution $p_i(T)=Z\,\frac{g_i}{g_0}\,e^{-\frac{\Delta E_i}{k T}}$.
Here $g_i$ ($g_0$) is the degeneracy factor of the state $|i\rangle$ (ground state),
$\Delta E_i$ the energy difference between $|i\rangle$ and the ground state, $k$ the Boltzmann constant
and $Z$ the normalization constant.

In our calculations we consider laterally coupled InGaAs/GaAs QDs.
We shall use parameters which, for long interdot distances, fit 
the emission spectrum observed in single QDs fabricated upon GaAs 
nanohole templates.\cite{AlonsoJCG,AlenXXX}
Effective masses are $m_e^*=0.0324$ and $m_h^*=0.435$, as corresponding
to In$_{0.75}$Ga$_{0.25}$As, and the dielectric constant $\epsilon=12.3$.\cite{Semi_Pam_Book}
The electron confinement frequencies are $\hbar \omega_{e,L}=35$ meV and
$\hbar \omega_{e,R}=34$ meV, so that the right QD is slightly bigger 
than the left QD. This is the situation found in some real samples.\cite{MunozJPCS,AlenXXX}
Hole confinement frequencies are taken such that $l_h=0.59\,l_e$, where
$l_i$ is the characteristic length, $l_i=\sqrt{\hbar / m_i^* \omega_i}$.
This means that holes are more confined than electrons, and their tunneling
will be negligible. Similar $l_h/l_e$ ratios have been estimated in recent
experiments with InGaAs QDs.\cite{TsaiPRL}
The bandwidth of the Lorentzian function in our spectral simulations 
is 0.05 meV. For the many-body calculation, we project Hamiltonian 
(\ref{Hfull}) onto all the single-particle configurations which can
be obtained by combining the 24 lowest-energy electron and hole
spin-orbitals. The origin of electron (hole) single-particle energies 
is taken at the edge of the conduction (valence) band, disregarding 
the $z$-confinement energy, which is constant for all the states.\\

\section{Results and Discussion}

\subsection{Exciton and negative trion}

\begin{figure}[h!]
\includegraphics[width=0.45\textwidth]{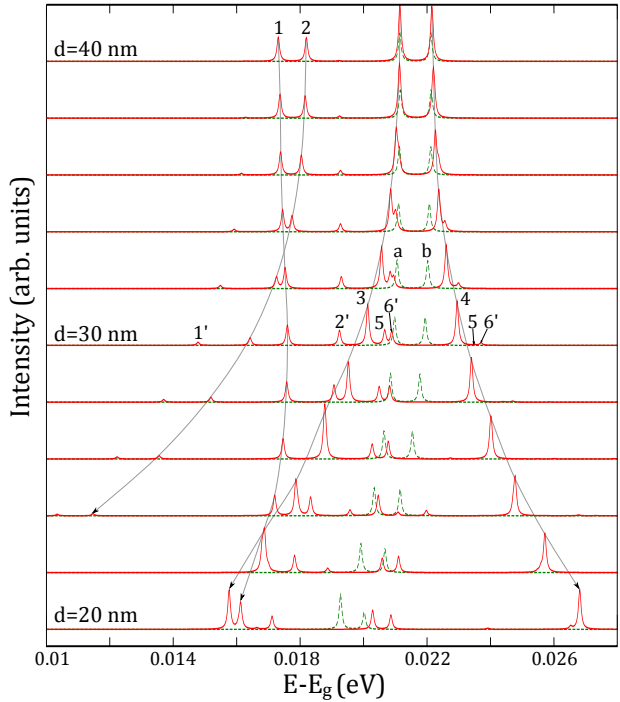}
\caption{(Color online). Emission spectrum of negative trions (solid red lines) and
neutral excitons (dashed green lines) for different interdot distances. 
Grey lines are a guide to the eyes. See text for the meaning of labels.}\label{fig:tot}
\end{figure}

Figure \ref{fig:tot} shows the emission spectrum of the QDM as
a function of the interdot distance $d$, from $d=40$ nm (isolated
QDs) to $d=20$ nm (strongly coupled QDs). Solid red and dashed green lines 
are used for negative trions ($X^-$) and neutral excitons ($X^0$), 
respectively. Here we assume an out-of-equilibrium system. 
In this way, we simulate optically active transitions from 
ground and excited states of both QDs, as normally
observed in the photoluminescence spectra of vertical~\cite{KrennerPRL,KrennerPRL2,StinaffSCI}
and lateral~\cite{MunozJPCS,AlenXXX,Zhou_arxiv} QDMs. 
The most remarkable result of Fig.~\ref{fig:tot} is that the $X^-$
and $X^0$ resonances follow very different evolutions as the QDs
are brought closer together. To understand the different behavior,
we need to analyze each excitonic complex in detail.

\begin{figure}[h!]
\includegraphics[width=0.45\textwidth]{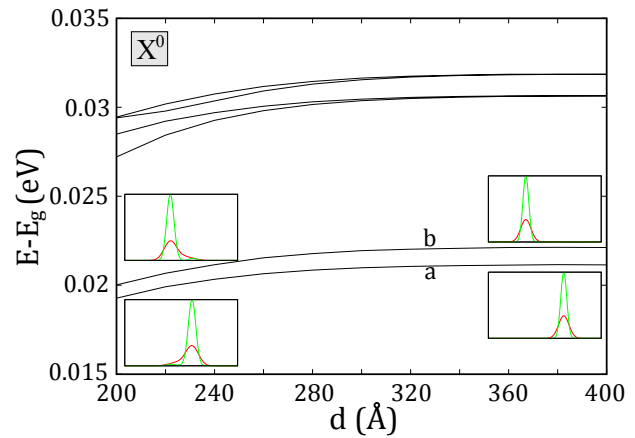}
\caption{(Color online). Energy of the lowest exciton states as a function of the 
interdot distance. The insets show the charge density of the electron (red line)
and hole (green line) within the exciton complex for the optically active states,
at $d=40$ nm (right) and $d=20$ nm (left).}\label{fig:X0}
\end{figure}

In the limit of the uncoupled QDM, $d=40$ nm, $X^0$ shows two resonances
of similar intensity split by $\sim1$ meV. 
As $d$ is reduced, the two resonances are slowly redshifted, 
but the energy splitting remains approximately constant. 
This behavior can be understood from the exciton states,
plotted in Fig.~\ref{fig:X0}. The emission resonances correspond to
the two lowest $X^0$ states.  At $d=40$ nm, these are the
direct excitons, with the electron-hole pair in the big 
(low-energy resonance) and small (high-energy resonance) QD, see
right-side insets in the figure.
Indirect excitons (electron and hole in opposite QDs) are
much higher in energy and optically dark because the electron-hole
overlap is negligible.
As the interdot distance decreases, tunnel coupling allows
electrons and holes to start hybridizing and form incipient
molecular orbitals, even if the QDs are not exactly resonant.
The lowest $X^0$ state corresponds to bonding (nodeless) 
electron and hole orbitals, so it is redshifted.
The first excited state corresponds to bonding electron but
antibonding hole orbital (note the presence of a node in the
green line of the upper-left inset). Yet, this makes little
difference because hole tunneling is negligible. 
As a result, the two states are redshifted by a similar amount.

The $X^0$ emission spectrum shown in Fig.~\ref{fig:tot} differs from 
previous calculations for asymmetric lateral QDMs, where the two lowest $X^0$
states were at some point blueshifted by Coulomb coupling.\cite{BairaJAP} 
This is due to the different confining potentials in the models. 
In Ref.~\onlinecite{BairaJAP}, the interdot barrier was a step-like 
potential. Instead, we use a parabolic potential such that, with
decreasing $d$, not only the QDs are closer but also the barrier
is lower. For this reason, tunnel coupling effects rapidly overcome
Coulomb coupling in our system. Real QDMs are lens-shaped QDs with
some degree of lateral contact.\cite{AlonsoJCG} As the dots are
brought closer, the height in the contact region increases.
We then believe our model is more appropriate.

Next we investigate the $X^-$ spectrum. As can be seen in Fig.~\ref{fig:tot},
at $d=40$ nm there are four resonances associated with $X^-$ states. 
Two resonances --(1) and (2)-- are a few meV below those of $X^0$, while two are degenerate with the $X^0$ ones.
As $d$ is reduced, several features show up which differ from the
simple behavior of neutral excitons: (i) the number of resonances
increases (up to 10); (ii) the two low-energy resonances invert
their ordering; (iii) the degeneracy between the high-energy $X^-$
and the $X^0$ resonances is lifted. Significantly, most of this occurs
even before the neutral exciton starts feeling tunnel coupling
(see e.g. $d=34$ nm). 

\begin{figure}[h!]
\includegraphics[width=0.45\textwidth]{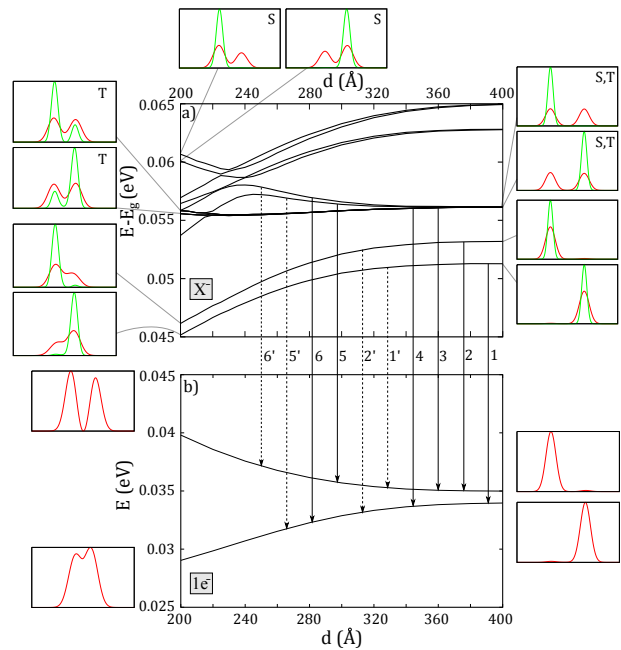}
\caption{(Color online). Energy of the lowest negative trion states (a) and electron
states (b) as a function of the interdot distance. The lateral insets show the 
charge density of the electron (red line) and hole (green line) 
at $d=40$ nm (right) and $d=20$ nm (left). Vertical arrows indicate the
transitions observed in Fig.~\ref{fig:tot}.}\label{fig:X-}
\end{figure}

To understand the $X^-$ emission spectrum, we study the energy levels 
of the initial ($X^-$) and final ($1e^-$) states, which are plotted in
Fig.~\ref{fig:X-}. The electron, Fig.~\ref{fig:X-}(b), displays a 
simple dissociation spectrum typical of diatomic heteronuclear molecules. 
At $d=40$ nm the electron is in one of the two QDs (see right insets). 
With decreasing $d$, it forms bonding and antibonding (noded) 
molecular orbitals (see left insets).
As for the trion, Fig.~\ref{fig:X-}(a), the two lowest-energy states 
correspond to the so-called direct trion states. Their behavior is the
same as that of $X^0$.  Namely, at $d=40$ nm all the carriers are 
localized inside one of the QDs, and they are redshifted with 
decreasing $d$ owing to electron hybridization (see insets).
As indicated by the vertical arrows in Fig.~\ref{fig:X-},
resonance 1 (2) in Fig.~\ref{fig:tot} originates in the transition
from the direct $X^-$ state in the right (left) QD to the electron
state in the same QD. The energy splitting between the 
two initial states is larger than that between the final states.
As a result, transition (1) is lower in energy.
However, as $d$ decreases the situation reverses. 
Transition (1) involves bonding initial and final states,
whose energetic stabilizations due to tunnel-coupling compensate each other.
As a result, the energy of the transition is little sensitive 
to the coupling (see Fig.~\ref{fig:tot}).
In contrast, transition (2) involves a bonding initial state 
but an antibonding final state, so its energy rapidly decreases.
This is also seen in Fig.~\ref{fig:tot} and leads to a reversal
of resonances (1) and (2) at $d\sim 32$ nm.
What is more, because of the QDM asymmetry, transitions between
bonding and antibonding states are not symmetry-forbidden.
Then, when electron hybridization takes place new transitions
start developing, (1') and (2').

Resonances (3),(4),(5) and (6) in the $X^-$ spectrum originate in a set
of excited trion states with energy $\sim 0.055$ eV at $d=40$ nm. 
These are the so-called indirect trions, where
one electron sits in a QD and the remaining electron-hole
pair sits in the opposite QD (see insets in Fig.~\ref{fig:X-}).
When the electron-hole pair recombines, the electron in the
opposite QD remains. At $d=40$ nm, the Coulomb interaction between 
the two QDs is so weak that the 
indirect trion resonances have the same energy as those of $X^0$.
With decreasing $d$, however, the behavior of indirect trions 
becomes noticeably different from that of neutral excitons. 
The indirect trion states split into two groups. 
One group is formed by states where the two electrons form a 
singlet spin configuration. These states increase in energy
and eventually anticross with higher states.
This destabilization is mainly caused by the gradual enhancement
of the antibonding character. 
In contrast, the group of states with electrons forming a triplet 
remains almost insensitive to tunnel-coupling due to the spin-blockade.
This effect is analogous to that observed in vertical QDMs under 
electric fields, where the Pauli exclusion principle prevents triplet 
states from tunneling.\cite{KrennerPRL}
 The characteristic evolution of indirect trions vs.~$d$ is responsible
for the lifting of the degeneracy with $X^0$ resonances observed in Fig.~\ref{fig:tot}.
Last, as in the case of direct trions, the QDM asymmetry enables
transitions from indirect trions to either bonding or antibonding electron states, 
which gives rise to new resonances when electrons hybridize, (5') and (6').

The different evolution of $X^-$ and $X^0$ resonances in the emission spectrum 
of QDMs can be used to elucidate the presence of molecular coupling.
The appearance of additional trion resonances should be a clear signature,
but they may be too weak for typical interdot distances ($d=30-40$ nm
in Ref.~\onlinecite{AlonsoJCG}). A more evident signature should be the splitting
between $X^0$ and indirect $X^-$ resonances, which involves intense optical resonances 
and is visible for long interdot distances ($d\sim36$ nm in Fig.~\ref{fig:tot}).
Finally, the reversal of resonances (1) and (2) may also be noticed in experiments. 
For $d<34$ nm, this implies that trion emission from the bigger QD occurs
at higher energy than that from the smaller QD, contrary to the ordering of $X^0$ resonances. 
As a matter of fact, this anomalous spectral ordering has been recently found  in
photoluminescence measurements of lateral QDMs.\cite{AlenXXX} 

\subsection{Positive trion}

\begin{figure}[h]
\includegraphics[width=0.45\textwidth]{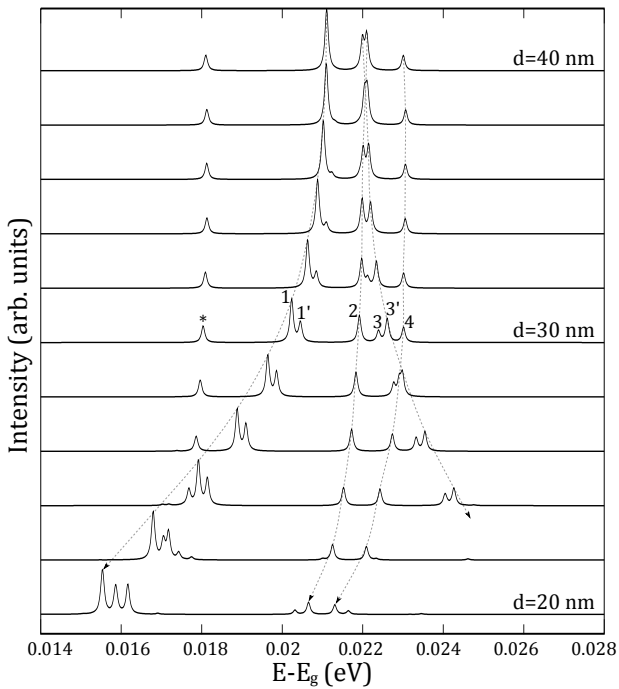}
\caption{Emission spectrum of positive trions for different interdot distances. Dashed lines
are a guide to the eyes.}\label{PLX+}
\end{figure}

We next study the emission spectrum of the positive trion ($X^+$). In this case, assuming
equal population for all the initial states would lead to a complex spectrum with many resonances
involving highly-excited initial and final states. Since we are interested in the main
transitions only, the emission spectrum plotted in Fig.~\ref{PLX+} has been computed 
assuming thermal equilibrium at $T=80$ K.
At this temperature, the $X^+$ states with significant population are essentially the same as previously
studied for $X^-$. 
The only remaining signature of transitions between excited states is the low-energy resonance 
labeled with an asterisk in Fig.~\ref{PLX+}, which will not be studied here.

\begin{figure}[h]
\includegraphics[width=0.45\textwidth]{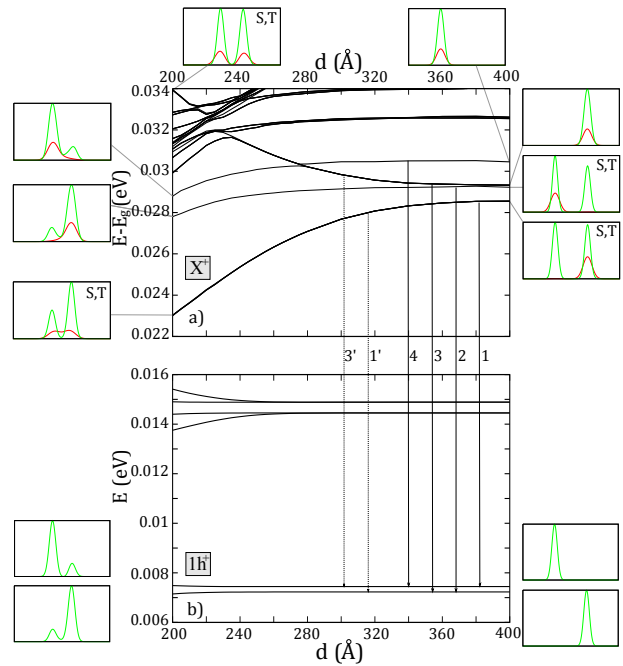}
\caption{(Color online). Energy of the lowest positive trion states (a) and hole states (b) as a function
of the interdot distance. The lateral insets show the charge density of the electron (red line) and the hole
(green line) at $d=40$ nm (right) and $d=20$ nm (left). Vertical arrows indicate the transitions observed
in~\ref{PLX+}.}
\label{X+}
\end{figure}

Comparing Figs.~\ref{PLX+} and~\ref{fig:tot} one can see that the $X^+$ emission pattern differs 
considerably from the $X^-$ one. 
To analyze the emission spectrum, in Fig.~\ref{X+} we represent the initial ($X^+$) and final ($1h^+$) 
states as a function of the interdot distance.  The first important difference with respect to $X^-$ 
can be seen in the final state, Fig.~\ref{X+}(b). Unlike for electrons, the two lowest
hole states remain almost unaltered throughout the entire range of $d$, due to the 
negligible tunneling. This means that the evolution of the $X^+$ emission spectrum can be 
entirely interpreted from the initial $X^+$ states. 
The second important difference is in the $X^+$ states, plotted in Fig.~\ref{X+}(a). 
There we find two direct and two indirect states, as in the $X^-$ case, but now their
energy ordering is reversed. At $d=40$ nm, the two lowest states are indirect trions, with one
hole in each QD and the electron in either the left or right QD (see insets).
We note that indirect $X^+$ ground states have also been obtained in 3D atomistic calculations~\cite{PengPRB2}, 
and are due the larger magnitude of the hole-hole Coulomb repulsion as compared to 
the electron-hole attraction, i.e.~ $\lvert V_{hh}\rvert > \lvert V_{eh} \rvert$. 
The situation is the opposite to that of $X^-$, where $\lvert V_{ee} \rvert < \lvert V_{eh} \rvert$, 
which favors direct trion configurations. 

The resonances labeled as (2) and (4) in Fig.~\ref{PLX+} originate in the recombination of the direct
trions. As the interdot distance is reduced they redshift slowly. 
 The evolution of the indirect trion resonances, (1) and (3), is however drastic. 
The two resonances become rapidly split with decreasing $d$, reflecting the strong
splitting of indirect trions in Fig.~\ref{X+}(a).
Such a large splitting follows from the strong tunneling of indirect $X^+$.
Unlike for $X^0$, the electron in indirect $X^+$ can tunnel back and forth without
losing the strong on-site Coulomb attraction $V_{eh}$. This clearly favors molecular coupling.\cite{exclusivo}
Furthermore, the electron delocalization over the entire QDM enables effective recombination with both holes, 
leading to the appearance of additional resonances,
(1') and (3'), at relatively long interdot distance (see e.g., $d=36$ nm).

\subsection{Low temperature spectra}

\begin{figure}[h!]
\includegraphics[width=0.45\textwidth]{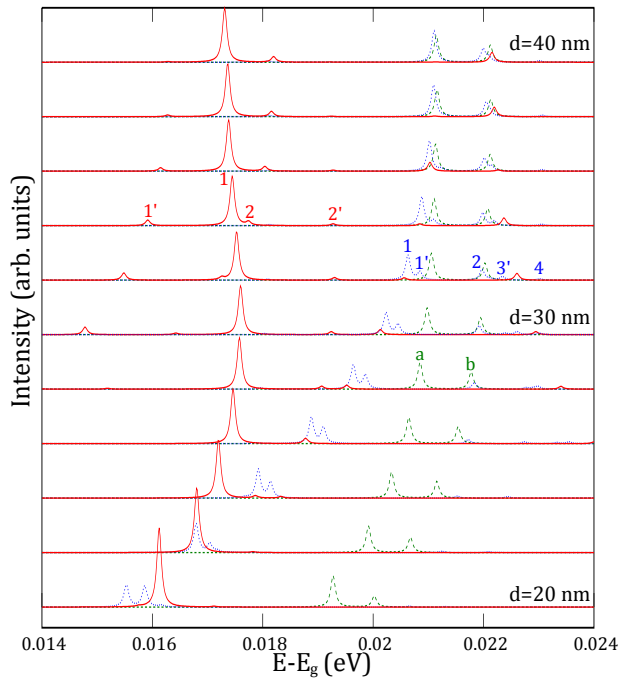}
\caption{(Color online). Emission spectrum of negative trions (solid red lines), 
neutral excitons (dashed green line) and positive trions (dotted blue lines) 
for different interdot distances.}\label{fig:T25}
\end{figure}

In this section, we re-examine the emission spectra assuming thermal
equilibrium at low temperature. This greatly simplifies the spectrum, 
as only populated initial states contribute to the spectrum. 
We consider $T=25$ K, which is enough for both QDs to have finite occupation.
Fig.~\ref{fig:T25} shows the emission spectrum for $X^-$ (solid red lines),
$X^0$ (dashed green lines) and $X^+$ (dotted blue lines).
The evolution of $X^0$ and $X^-$ ($X^+$) resonances is essentially that of the direct 
(indirect) species discussed above. Under these conditions, the only signatures of
coupling remaining are: (i) the appearance of weak $X^-$ resonances, labeled 
in red as (1') and (2'), and (ii) the presence of strong low-energy $X^+$ resonances.
These transitions arise from the ground (indirect) $X^+$ state, which soon split 
into (1) and (1') and undergo a rapid redshift with decreasing $d$.
Thus, at $d=40$ nm we start with a typical spectrum of isolated QDs, where $X^0$ and $X^+$ resonances are
close in energy, while $X^-$ resonances are a few meV below.\cite{DalgarnoPRB,ClimentePRB}
As $d$ is reduced, the $X^+$ resonances redshift and approach those of $X^-$.

\section{Conclusions}

We have compared the emission spectrum of $X^0$, $X^-$ and $X^+$ in asymmetric
lateral QDMs as a function of the interdot distance. We have shown that tunneling of
trions is stronger than that of neutral excitons. This is because the net Coulomb
interactions the electron has to overcome in order to tunnel are smaller.
As a result, $X^{\pm}$ display signatures 
of molecular coupling at longer interdot distances than $X^0$. 
The signatures include:
(i) the appearance of additional optical resonances;
(ii) the inversion of the ordering of direct trion resonances corresponding 
to the big and small QDs;
(iii) the lifting of the degeneracy between indirect trion and neutral exciton
resonances;
(iv) a pronounced redshift of the indirect $X^+$ resonance, which approaches that of $X^-$. 
We note that signature (ii) has been actually observed in recent photoluminescence
experiments.\cite{AlenXXX}
We have also shown that $X^-$ and $X^+$ develop different emission patterns. 

These results are particularly valuable in the view of recent experimental
progress to control the charge of excitonic complexes in QDMs.\cite{MunozJPCS,AlenXXX,Zhou_arxiv}
It has been recently shown that trions are more sensitive than excitons to electric 
fields, owing to their net electric charge.\cite{PengPRB2} This, together with
our results, suggests that current attempts to utilize lateral QDMs should focus
on trion species rather than neutral excitons. In both positive and negative
trions electron tunneling can be efficient. Yet, positive trions might be less
suited because the spin of holes is very sensitive to the in-plane confinement
anisotropy.\cite{KoudinovPRB,DotyPRB} In a lateral QDM, the $X^+$ ground state
contains one hole in each QD. The holes then feel the mutual Coulomb repulsion 
as a source of anisotropic confinement, which translates into spin mixing. 
This could hamper the implementation of spin-based quantum information protocols.

\begin{acknowledgments}
We thank B. Al\'en, J. Mart\'{\i}nez-Pastor and G. Mu\~{n}oz-Matutano for fruitful discussions.
Support from MCINN project CTQ2008-03344, UJI-Bancaixa project P1-1A2009-03, a Generalitat Valenciana FPI grant (MR)
and the Ramon y Cajal program (JIC) is acknowledged.
\end{acknowledgments}


\begin{thebibliography}{35}
\bibitem{exclusivo} This behavior is exclusive of $X^+$, because for indirect $X^-$ the hole tunneling is negligible.
\end{thebibliography}
\end{document}